\documentclass[journal]{IEEEtran}
\usepackage{amsmath,epsfig}

\usepackage{graphicx}
\usepackage{caption}
\usepackage{subfig}
\usepackage{amsmath}
\usepackage{amssymb}
\usepackage{epsfig}
\usepackage{cite}
\usepackage{color}
\usepackage{balance}
\usepackage{amsmath}
\usepackage{amsfonts} 
\usepackage{graphicx}
\usepackage{pdfpages}
\usepackage[T1]{fontenc} 
\usepackage{array}
\usepackage{url}

\usepackage{color}
\usepackage{wrapfig}
\usepackage{lipsum}
\usepackage[hidelinks]{hyperref}
\usepackage{multirow}
\usepackage{tabularx}
\usepackage{gensymb} 
\usepackage{breqn}
\usepackage{array}
\usepackage{booktabs}
\usepackage{textcomp}
 
\begin{document}
\title{A Data-Driven Approach\\for Accurate Rainfall Prediction}

\author{Shilpa~Manandhar, \IEEEmembership{Student~Member,~IEEE,}
        Soumyabrata~Dev, \IEEEmembership{Member,~IEEE,}
        Yee~Hui~Lee, \IEEEmembership{Senior~Member,~IEEE,}
        Yu~Song~Meng, \IEEEmembership{Member,~IEEE,}
        and Stefan~Winkler,~\IEEEmembership{Fellow,~IEEE
        }
\thanks{Manuscript received $16^{th}$ Aug, $2018$, revised $10^{th}$ Dec, $2018$, accepted $18^{th}$ June, $2019$.  This work was supported in part by the Defence Science and Technology Agency, Singapore.}
\thanks{S. Manandhar and Y. H. Lee are with the School of Electrical and Electronic Engineering, Nanyang Technological University, Singapore.}
\thanks{S. Dev is with ADAPT SFI Research Centre, Trinity College Dublin, Ireland.}
\thanks{Y. S. Meng is with National Metrology Centre, Agency for Science, Technology and Research (A$^{*}$STAR), Singapore.}
\thanks{S. Winkler is with School of Computing, National University of Singapore.}
}

\maketitle

\begin{abstract}
In recent years, there has been growing interest in using Precipitable Water Vapor (PWV) derived from Global Positioning System (GPS) signal delays to predict rainfall. However, the occurrence of rainfall is dependent on a myriad of atmospheric parameters. This paper proposes a systematic approach to analyze various parameters that affect precipitation in the atmosphere. Different ground-based weather features like \emph{Temperature}, \emph{Relative Humidity}, \emph{Dew Point}, \emph{Solar Radiation}, \emph{PWV} along with  \emph{Seasonal} and  \emph{Diurnal} variables are identified, and a detailed feature correlation study is presented. While all features play a significant role in rainfall \emph{classification},  only a few of them, such as \emph{PWV}, \emph{Solar Radiation}, \emph{Seasonal} and \emph{Diurnal} features, stand out for rainfall \emph{prediction}. Based on these findings, an optimum set of features are used in a data-driven machine learning algorithm for rainfall prediction. The experimental evaluation using a four-year (2012-2015) database shows a true detection rate of 80.4\%, a false alarm rate of 20.3\%, and an overall accuracy of 79.6\%. Compared to the existing literature, our method significantly reduces the false alarm rates. 

\end{abstract}

\begin{IEEEkeywords}
precipitation, PWV, remote sensing, machine learning.
\end{IEEEkeywords}

\IEEEpeerreviewmaketitle

\section{Introduction}
\label{sec:intro}

\IEEEPARstart{R}{ainfall} initiation is a dynamic process and is influenced by a myriad of atmospheric parameters. The water vapor content of the atmosphere is one such important parameter. It is generally explained in terms of Precipitable Water Vapor (PWV) -- a measure of the total water vapor stored in a column of the atmosphere. It is an important indicator of water vapor climatology in the lower troposphere \cite{Jin_2007,Jin_2008}. Nowadays, Global Positioning System (GPS) signal delay is extensively being used to estimate  PWV, because GPS meteorology offers improved spatial and temporal resolutions for water vapor variations 
compared to other existing techniques like radiosondes, microwave radiometers and satellite-based instruments. Radiosondes are generally launched only twice a day and are not released during severe weather events \cite{Wang_2017}. Microwave radiometers have sparse station distribution as the instrument cost is high. They are of limited value in climate studies, particularly in predicting and tracking heavy rainfall cases, because radiometers can provide reliable PWV readings only under no-rain conditions \cite{Yeh_2018}. Similarly, satellite-based PWV retrieval has poor temporal resolution. Compared to these technologies, GPS provides good spatio-temporal resolution and is suitable for all weather conditions.

With the rapid deployment of GPS monitoring stations at local, regional, and global scales, there has been a renewed interest in using GPS-derived PWV for the prediction of rainfall. Seco et al.\ \cite{Seco_2012}, Manandhar et al.\ \cite{IGARSS_PWV} and Shi et al.\ \cite{Junbo} have presented some cases of severe- and moderate- rainfall events 
to indicate the feasibility of GPS-derived PWV for rainfall monitoring and forecasting. GPS-derived PWV was also used for studying a heavy rainfall event in Japan \cite{Shoji_2009} and flash flood events in France \cite{Brenot_2006}. However, Shi et al.\ \cite{Junbo} concluded that high PWV does not necessarily indicate the occurrence of rainfall; external dynamic factors also play a role in triggering a rain event. In \cite{Benevides,Yibin(a),Yibin(b)}, PWV and its derivatives of $6$-hour duration are used in forecasting a rain event within the next $6$ hours. Results have shown a detection rate of above $80$\%, but the false alarm rates are also high at $60$-$70$\%. Therefore, it can be concluded that PWV values are a good indicator of rain, but other factors need to be considered to improve the accuracy of prediction of rainfall events. 

Li et al.\ \cite{Li_2005} suggested that meteorological parameters like temperature along with PWV could  be useful for prediction of rainfall. Similarly, Sharifi et al.\ \cite{sharifi} proposed to use the relative humidity anomaly along with the PWV anomaly values to improve the prediction accuracy of rainfall events. In the literature, there are only a few papers that combine different weather features and use a machine-learning based methodology 
for rainfall prediction~\cite{Luk_2000,Sawale_2013,Hernandez_2016}. Most of these focus on implementation and inter-comparison of different machine learning algorithms. However, for the meteorology community, it is more interesting to identify the important \emph{parameters} for rainfall prediction, their inter-dependency and their level of contribution in the prediction.

In this paper, a detailed study of different weather parameters is presented. Along with the different weather parameters, the seasonal and diurnal variables are also considered, which are generally neglected in most related studies. 
These weather parameters and seasonal factors are individually assessed for rainfall prediction. Those parameters that are important for rainfall prediction are identified, and a machine learning algorithm is implemented, which shows significant improvements in rainfall prediction accuracy as compared to the existing literature. 
   
The rest of this paper is organized as follows. Section II gives a brief overview of the weather station data and GPS data that are used in this paper. Section III identifies different weather parameters that can be useful for rainfall prediction. Section IV gives a detailed analysis of the interdependency of different weather parameters. Section V describes the mathematical tools and methods that are used. Section VI discusses the rainfall prediction results.. Finally, conclusions and future work are presented in Section VII.
 
\section{Meteorological Sensor Database and Data Processing}
In this section, we briefly describe the different parameters that are used in this paper, which include surface weather parameters and total column water vapor content.

\subsection{GPS-Derived Water Vapor Content}
PWV values (in mm) are derived from GPS signal delays. GPS signal delays, generally referred to as the zenith total delay (ZTD), can be broadly classified into zenith wet delay (ZWD) and zenith dry delay (or hydrostatic delay) (ZHD). Out of these delays, ZHD contributes about $90$\% of the total zenith delay and is dependent on the surface pressure, temperature and refractive index of the troposphere \cite{Elgered_1991}. In contrast, ZWD contributes very little to the total zenith delay and is a function of atmospheric water vapor profile and humidity. ZWD is used for the calculation of  PWV  as ZWD is related to the moisture profile of the atmosphere. 

There are different empirical models that can be used to derive  ZHD, such as the Saastamonien equation, VMF1 model or Static model. The Saastamonien equation uses pressure values to calculate the hydrostatic delays \cite{Jiang_2016}. The actual pressure and temperature values are not readily available for all the GPS stations. In such cases, pressure values derived from empirical models like GPT (Global Pressure Temperature) or GPT2 can be used \cite{LIU_ZHD_2017}. Alternatively, Vienna Mapping Function I (VMFI) model can be used, which provides ZHD values for different stations \cite{Leontiev_2017}. The VMF1 model derived hydrostatic delays are provided in the Global  Geodetic  Observing  System (GGOS) website \cite{GGOS_site} and are based on  meteorological data from Numerical Weather Models (NWMs). ZHD  can also be derived using a static model that is based on the station height only \cite{Astudillo_2018}. Static models are more appropriate for tropical stations, where temperature variations are minimal.

It is relatively difficult to calculate ZWD  as there are no empirical models. For this paper, the ZWD values are processed using GIPSY-OASIS software (GPS Inferred Positioning System Orbit Analysis Simulation Software package) and its recommended scripts \cite{NASA_Rinex}. The GIPSY processing was carried out using the default ZHD model of the software (static ZHD model) with an elevation cutoff angle of $10^{\circ}$ and the Niell mapping function.

Once the ZWD (\textit{$\delta$L$_w^{o}$}) values are estimated using the software, PWV is calculated using \ref{eq1}, as follows: 
\begin{equation}
	\mbox{PWV}=\frac{PI \cdot \delta L_w^{o}}{\rho_l},
    \label{eq1}
\end{equation}     
where $\rho_{l}$ is the density of liquid water ($1000$ kg$/m^{3}$). $PI$ is a dimensionless factor determined by using Eq.~(\ref{eq:PI}), which was derived using radiosonde data from $174$ stations in our previous paper \cite{shilpaPI}: 
\begin{dmath}
	PI=[-\textrm{sgn}(L_{a}) \cdot 1.7\cdot 10^{-5} |L_{a}|^{h_{fac}}-0.0001] \cdot \cos\frac{2\pi(DoY-28)}{365.25}+0.165-1.7\cdot 10^{-5}|L_{a}|^{1.65}+f,
    \label{eq:PI}
\end{dmath}
where $L_{a}$ is the latitude, $DoY$ is day-of-year, $h_{fac}=1.48$ for stations from northern hemisphere and $1.25$ for stations from southern hemisphere. $f=-2.38\cdot 10^{-6}H$, where \textit{H} is the station height, which can be ignored for stations below $1000$m. 

Here, the PWV values are calculated for tropical GPS stations; station ID: NTUS at ($1.30$$^{\circ}$N, $103.68$$^{\circ}$E) with station height above sea level $\approx 79$m,  and SNUS at ($1.29$$^{\circ}$N, $103.77$$^{\circ}$E) with station height  above sea level $\approx 63$m, both located in Singapore. NTUS is under IGS network, and SNUS is part of the Singapore  Satellite Positioning  Reference Network (SiReNT) under Singapore Land Authority (SLA) \cite{SiReNT}. PWV can then be calculated for NTUS and SNUS using Eq.~(\ref{eq1})-(\ref{eq:PI}) with respective values for $L_{a}$, $h_{fac}$, $H$ and $DoY$. The resulting PWV values have a temporal resolution of $5$ minutes. In this paper, $4$ years ($2012$-$2015$) of PWV data from NTUS and $1$ year ($2016$) of PWV data from SNUS are used.

\subsection{Weather Station Data}
In addition to the PWV values, we use different surface weather parameters like temperature ($T$, $^{\circ}$C), relative humidity ($RH$,\%), dew point temperature ($DPT$, $^{\circ}$C) and solar radiation ($SR$, W/m$^{2}$). The surface weather parameters along with the rainfall rates (mm/hr) are recorded by the weather stations co-located with the GPS stations. The weather station at NTUS records weather data at an interval of $1$ minute and is maintained by our group. The weather station at SNUS records data at an interval of $5$ minutes and is available online \cite{NUS_WS}. $4$ years ($2012$-$2015$) of weather variables from NTUS and $1$ year ($2016$) of weather variables from SNUS are used. The GPS and weather station data from NTUS station are used in the proposal of the algorithm and data from SNUS station are used in benchmarking and validation.

In addition to the data from the weather station, we also use images from the ground-based sky cameras called Wide Angle High Resolution Sky Imaging System (WAHRSIS) co-located with NTUS station. This allows for a visual validation of the atmospheric conditions. 
In the following sections, the weather station parameters are sampled every $5$ minutes to match the GPS-PWV timings for the NTUS station.

\section{Approach \& Tools}
\label{sec:method}
In this section, we describe the different tools and techniques that have been implemented to use the various weather features for prediction of rain events.

\subsection{Supervised Machine Learning Technique}
For the purpose of rainfall prediction, the samples are labeled as either rain or no-rain.  Therefore, supervised machine learning techniques can be implemented for rainfall prediction. Artificial Neural Network (ANN) and Support Vector Machine (SVM) are the most commonly used supervised machine learning techniques for dealing with such geoscience-based problems~\cite{SVM_Lary}. We use SVM as it is effective and computationally efficient.

SVM is a parametric method that generates a hyperplane or a set of hyperplanes in the vector space by maximizing the margin between classifiers to the nearest neighbor data \cite{SVM_Soumya}. In this paper, we use SVM to classify rain and no-rain cases using different weather parameters as features. Consider a feature matrix $X$ of dimension $m \times n$, where $n$ is the number of features (weather variables like $T$, $DPT$, $RH$, etc.) and $m$ is the number of samples. Consider an output matrix $Y$, a column matrix of dimension $m \times 1$, where values can be either $1$ or $-1$, indicating rain or no-rain, respectively. SVM is trained with a data set of $i$ points represented by 
$(\vec{x_{1}},y_{1}), (\vec{x_{2}},y_{2}), \ldots, (\vec{x_{k}},y_{k})$, where $\vec{x_{i}}$ is the $i^{th}$
row of the feature matrix $X$ and $y_{i}$ is the $i^{th}$ coefficient of the output matrix $Y$. $k$ depends on the training set size. 

Based on the training samples, SVM generates a maximum-margin hyperplane that separates the group of points $\vec{x_{i}}$ for which $y_{i}=1$ (i.e.\ rain) from the group of points for which $y_{i}=-1$ (i.e.\ no-rain). This is done such that the distance between the hyperplane and the nearest point $\vec{x_{i}}$ from either group is maximized. After the SVM is trained, the remaining samples ($m-i$) are used in testing the model. The predicted output is then compared to the real observation data and evaluation metrics are calculated. 

\subsection{Evaluation Metrics} 
The performance of rainfall prediction methods are generally expressed in terms of true detection and false alarm rates~\cite{Yibin(a), Benevides}. 
Table \ref{table:confusionII} shows the confusion matrix, indicating all possible cases when the predicted output is compared to the real observation data. In the evaluation, the true positive ($TP$), true negative ($TN$), false positive ($FP$) and false negative ($FN$) samples are calculated. The true detection rate is defined as $TD = TP/(TP+FN)$, the false alarm rate is defined as $FA = FP/(TN+FP)$.  We also report accuracy, which is defined as $A=(TP+TN)/(TP+FN+TN+FP)$, and the missed detection rate of the algorithm, $MD=1-TD$.

\begin{table}[htb!]
\begin{center}
\caption{Confusion Matrix}
\begin{tabular}{|c|c|c|}
\hline
              & Predicted (No)        & Predicted (Yes)        \\
\hline
Actual (No)   & True Negative (TN)    & False Positive (FP)    \\
\hline
Actual (Yes)  & False Negative (FN)   & True Positive (TP)     \\
\hline
\end{tabular}
\label{table:confusionII}
\end{center}
\end{table}

\subsection{Downsampling Technique}
In the following, SVM is trained and tested using our weather database. 
The weather data have a temporal resolution of $1$ minute, therefore one year database generally includes a total of $365*1440$ data points. Out of these $525,600$ data points, there are far fewer data points with rain than without, because rain is a relatively rare event. For example in the year $2015$ for NTUS station there are $525,120$ valid data points. Out of these, there are only $5,017$ data points with rain, referred to as \textit{minority} cases, whereas there are $520,103$ data points without rain (\textit{majority} cases). The \textit{minority} to  \textit{majority} ratio here is nearly $1:104$, which poses the problem of a highly imbalanced dataset. 
Training a model with such skewed data (skewed towards non-rain data points) would result in a biased model, which is dominated by the characteristics of the \textit{majority} database \cite{Ruiz_2008, shilpa_IGARSS_svm}, and compromises the generalization ability of the algorithm \cite{Vonikakis_2016}. 

Therefore, we employ a downsampling technique that balances the number of positive and negative labels. We consider all the cases from the \textit{minority} scenario and the cases from \textit{majority} scenario are randomly chosen such that the \textit{minority} to \textit{majority} ratio is balanced. There is a general practice to make the ratio $1:1$, but other ratios can also be considered \cite{RandomDownSampling}.
All the prediction results presented in this paper are after implementing the downsampling technique. In tests without downsampling, the accuracy of the algorithm dropped and the confidence interval range increased significantly.

\section{Feature Identification \& Correlation}

\subsection{Features}
We identify five important weather features: $T$, $RH$, $SR$, $DPT$ and $PWV$. These weather features have inherent diurnal and seasonal properties, which can be helpful in rainfall prediction.

In addition, identify the seasonal and diurnal features for our proposed task.  In the tropical climate of Singapore, four main seasons are experienced -- North-East (NE) Monsoon from November to March, First-Inter (FI) Monsoon from April to May, South-West (SW) Monsoon from June to October, and Second-Inter (SI) Monsoon from October to November. The occurrence of different seasons changes slightly from year to year as reported in the yearly weather report~\cite{NEA}. The rainfall pattern shows some correlation with the seasons. We often experience late afternoon showers during the NE Monsoon. Sumatra squalls are experienced during pre-dawn to midday, and short-lived rainfall often takes place in the afternoon during the SW Monsoon. During inter-monsoon seasons, afternoon to early evening rain events are common~\cite{NEA,TGRS_RainNowcasting}. Therefore, we consider day-of-year ($DoY$) as a feature that takes the seasonal effect into consideration.

The diurnal characteristics of all the features can be clearly observed in the time series observation from Fig.~\ref{fig:TimeSeries}, which will be discussed in detail in Section \ref{sec:TimeSeries}. We consider hour-of-day ($HoD$) as a feature that takes the diurnal effect into consideration. 

\begin{figure}[htb!]
\begin{center}
\subfloat[]{%
\includegraphics[width=0.25\textwidth]{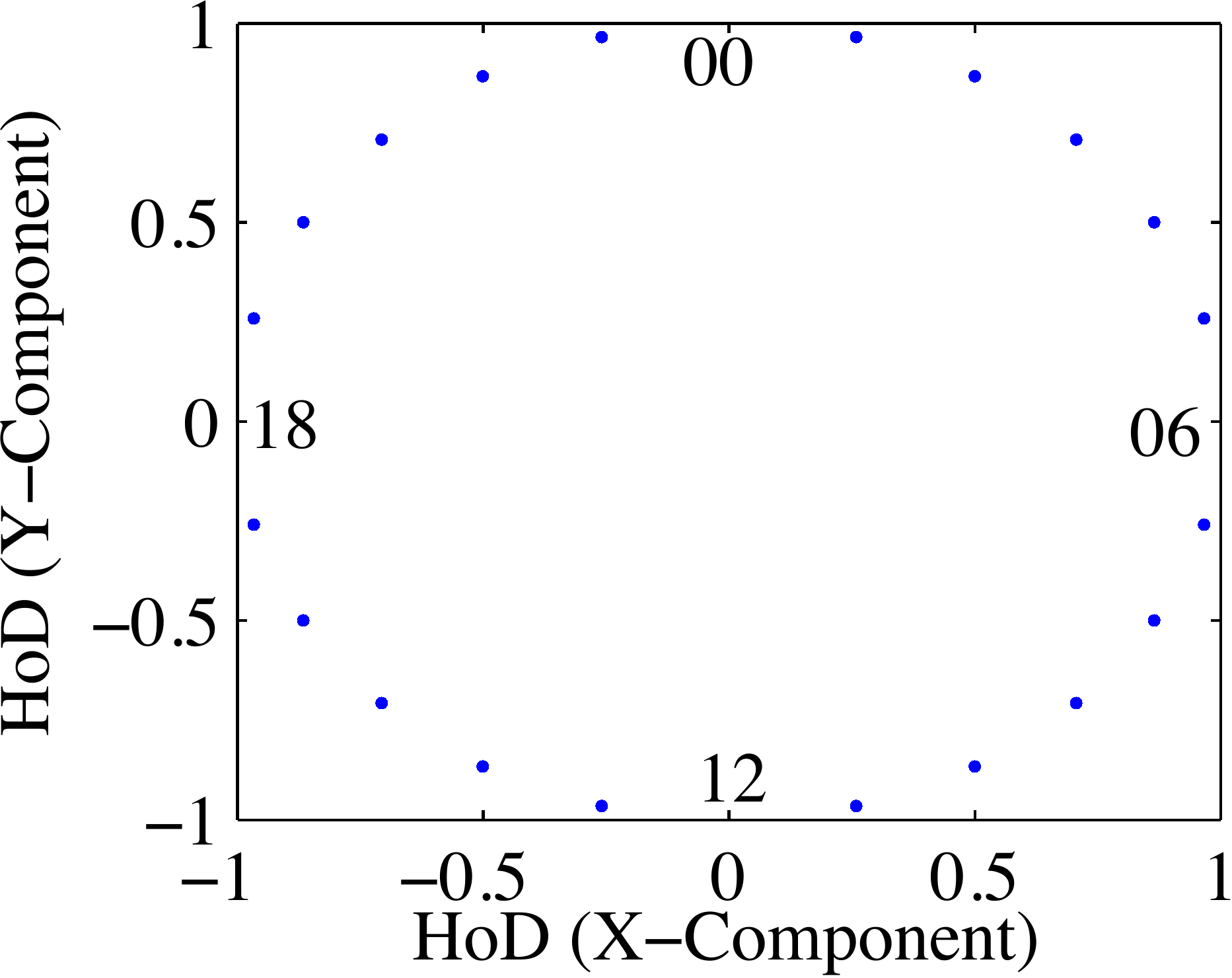}}
\subfloat[]{%
\includegraphics[width=0.25\textwidth]{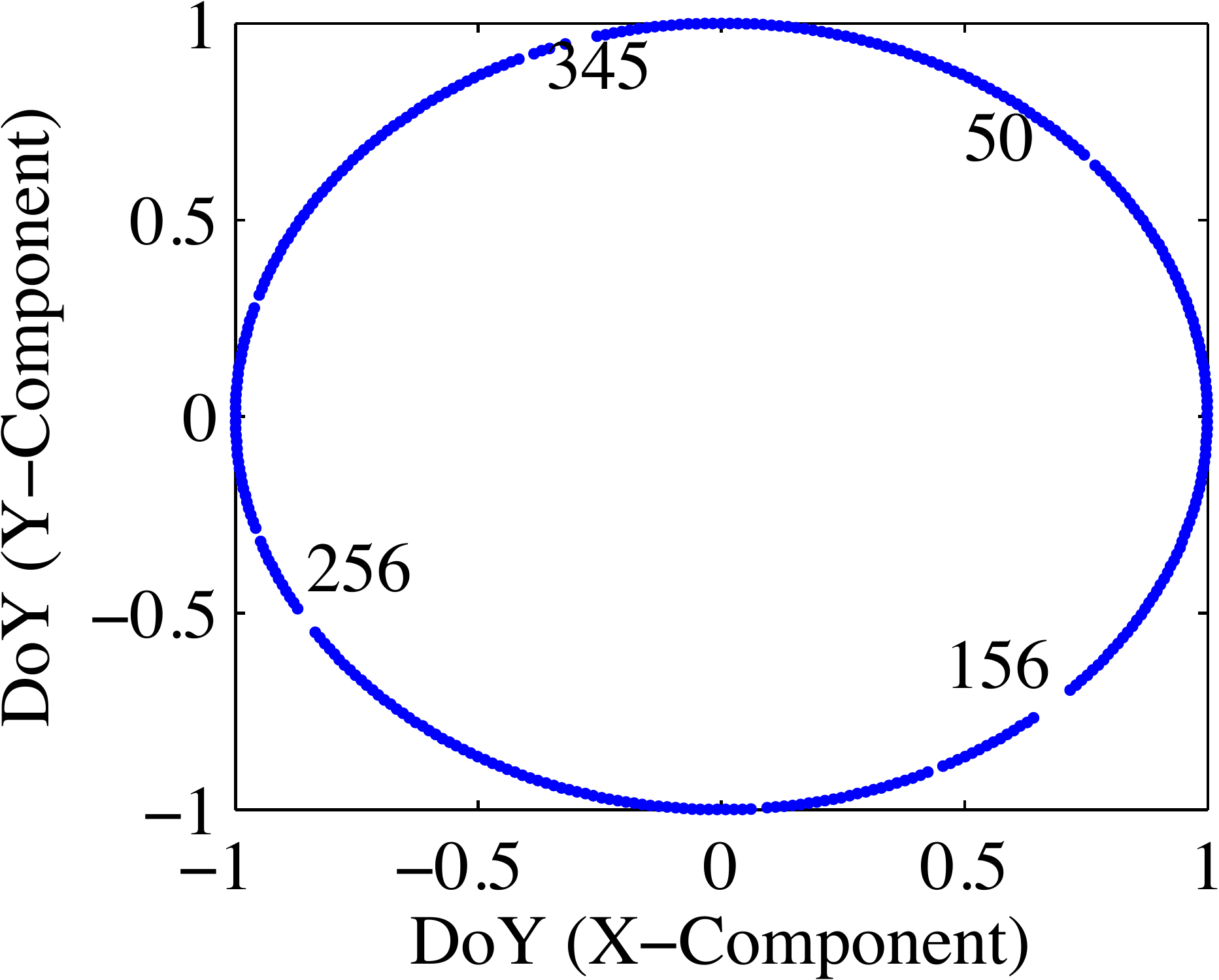}}
\caption{(a) X and Y components of $HoD$; (b) X and Y components of $DoY$. The numbers indicate the hour-of-day and day-of-year in plots (a) and (b) respectively.
\label{fig:cyclic}}
\end{center}
\end{figure}

For $HoD$ feature, the hour values reset after every $24$ hours, and for the $DoY$ feature the number of days resets after every $365$ days ($366$ for leap years). Thus, the $DoY$ and the $HoD$ features are both cyclic in nature, as the same values repeat after a specific period of time.
Therefore, each of these features $DoY$ and $HoD$ are expressed into its \emph{sine} and the \emph{cosine} components so that their cyclic properties are properly captured. Eq.~(\ref{eq.hx}) and (\ref{eq.hy}) are used for expressing the feature $HoD$ into its \emph{sine} and \emph{cosine} components; $HoD_{x}$ and $HoD_{y}$ respectively and are plotted as shown in Fig.~\ref{fig:cyclic}(a). The $DoY$ feature can be similarly expressed into its \emph{sine} and \emph{cosine} forms; $DoY_{x}$ and $DoY_{y}$ using Eq.~(\ref{eq.Dx}) and (\ref{eq.Dy}) respectively. Fig.~\ref{fig:cyclic}(b) shows the $DoY_{x}$ and $DoY_{y}$ values. 
\begin{equation}
	\mbox{$HoD_{x}$}=\cos \frac{2\pi \cdot HoD}{24}
    \label{eq.hx}
\end{equation} 
\begin{equation}
	\mbox{$HoD_{y}$}=\sin \frac{2\pi\cdot HoD}{24}
    \label{eq.hy}
\end{equation}
\begin{equation}
	\mbox{$DoY_{x}$}=\cos \frac{2\pi\cdot DoY}{365}
    \label{eq.Dx}
\end{equation} 
\begin{equation}
	\mbox{$DoY_{y}$}=\sin \frac{2\pi\cdot DoY}{365}
    \label{eq.Dy}
\end{equation}
In the rest of the paper, if any method includes the $HoD$ or $DoY$ features, their \emph{sine} and \emph{cosine} components are used.

\subsection{Feature Correlation}
It is important to analyze the correlation between different features, because if two features are perfectly correlated, the second feature will not provide any additional information, as it is already determined by the first~\cite{shilpa_IGARSS_feature}.

\begin{figure}[htb]
\begin{center}
\includegraphics[width=0.5\textwidth]{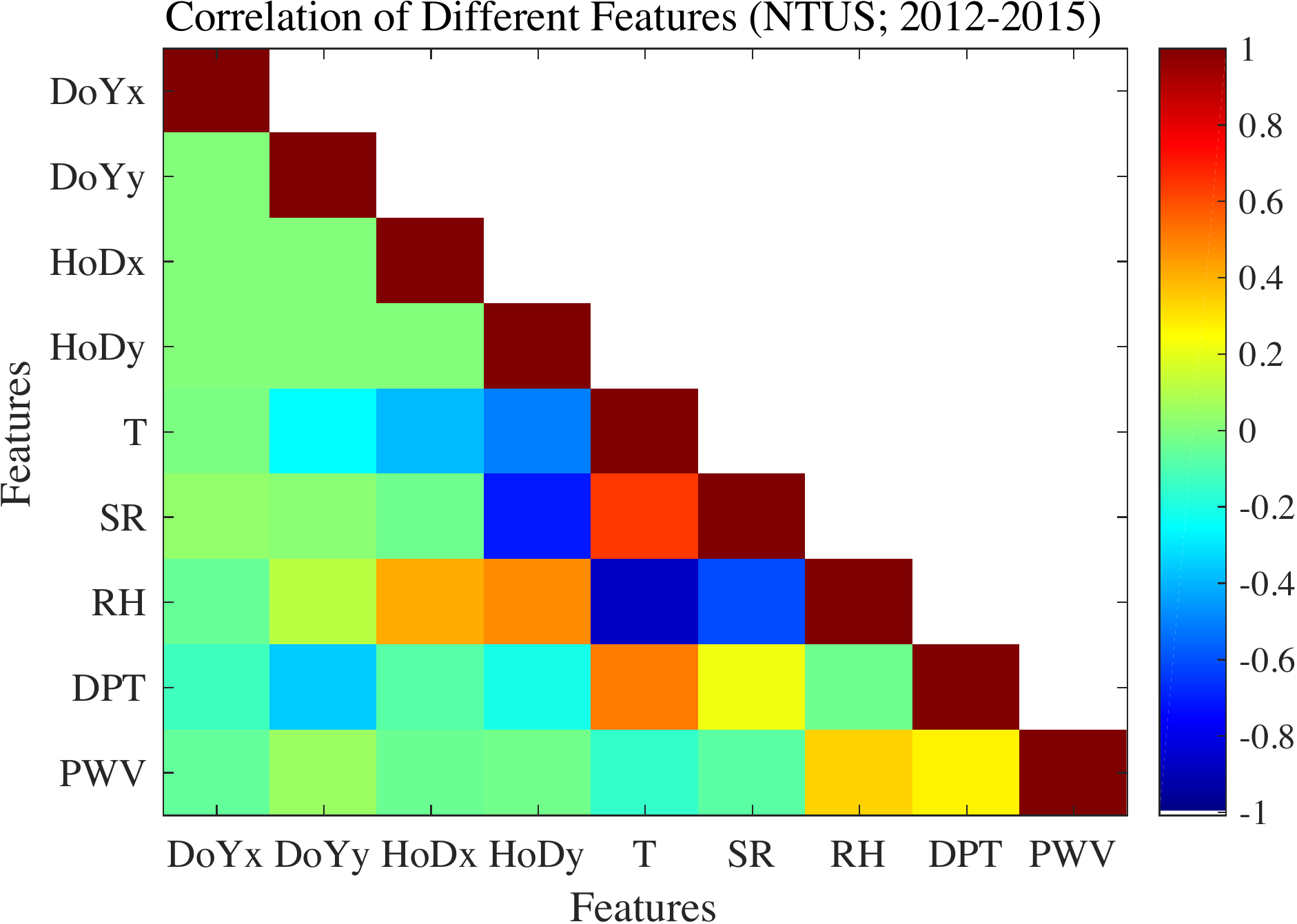}
\caption{Degree of correlation between various features (\emph{best viewed in color}).
\label{fig:cmat}}
\end{center}
\end{figure}

\begin{figure*}[htb!]
\centering
\includegraphics[width=1\textwidth]{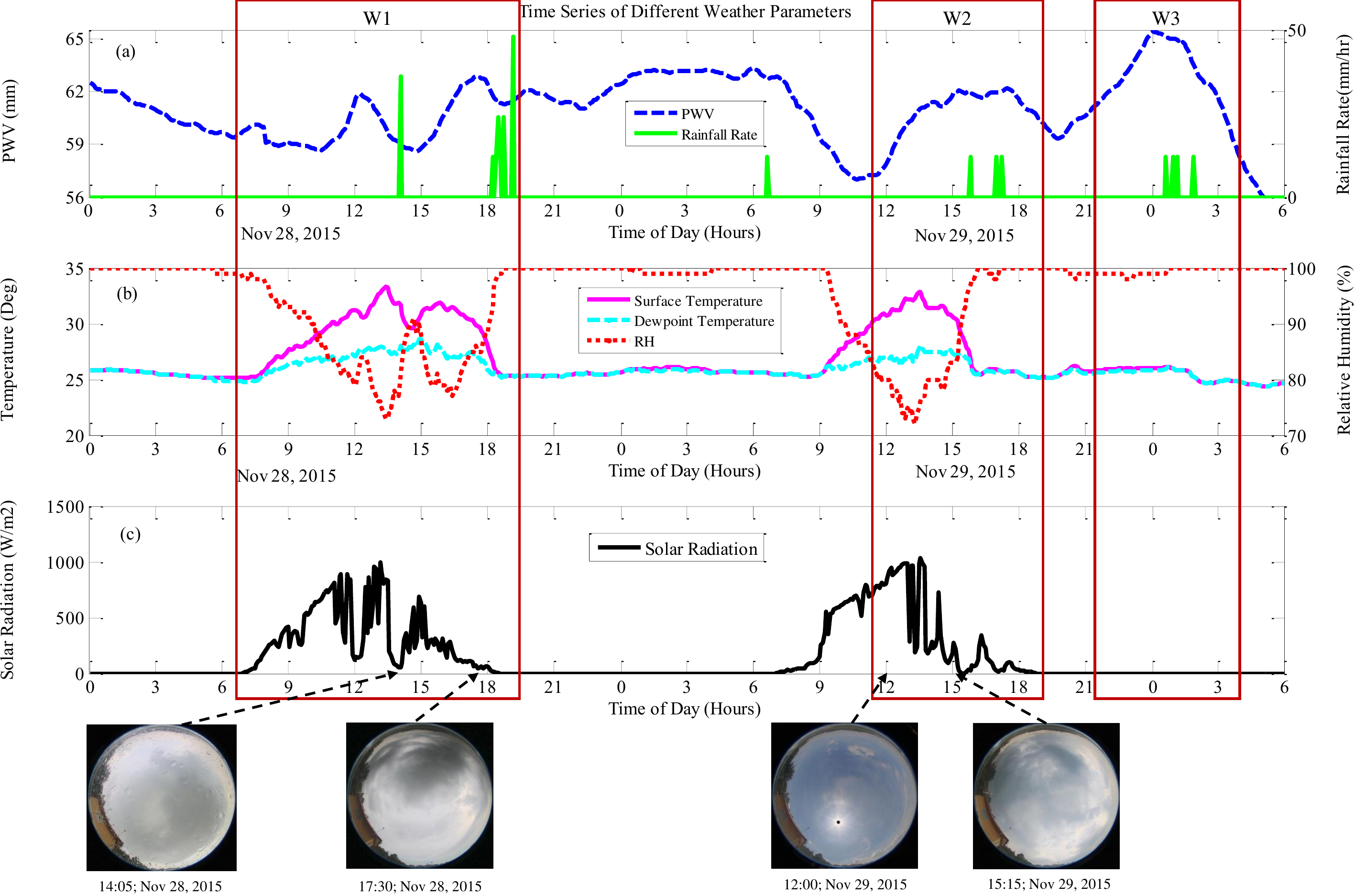}
\caption {Time series of different weather parameters for NTUS. (a) PWV (blue dashed line) in mm and Rainfall Rate (green solid line) in mm/hr; (b) Surface Temperature in $^{\circ}$C (solid magenta line), Dew point in $^{\circ}$C (dashed cyan line) and Relative humidity in\% (red dots); (c) Solar Radiation in W/m$^{2}$. The X-axis for all the subplots represents the time of day (local time, UTC+8). The sky images captured by WAHRSIS are also shown for specific times of the day. Windows W1, W2 highlight day time rain events, and W3 highlights a night time rain event (\emph{best viewed in color}). 
\label{fig:TimeSeries}}
\end{figure*}

The degree of correlation amongst all the features is shown in Fig.~\ref{fig:cmat}. Data from year $2012$-$2015$ of NTUS station are used in plotting the figure. Various observations can be made from the off-diagonal elements. $T$ has a good correlation with features $RH$, $SR$, and $DPT$. A high negative correlation coefficient of around $-0.9$ is observed between $T$ and $RH$, which is expected because when the air is warm, it can hold more water vapor, thus the saturation point increases and the relative humidity becomes lower. $T$ and $SR$, and $T$ and $DPT$ have positive correlation coefficients of $0.6$ and $0.5$. 
During the day time, as the sun rises, $SR$ and $T$ values both increase, while $RH$ decreases. The features $RH$ and $SR$ are negatively correlated with a correlation coefficient of around $-0.6$. This can be explained as $SR$ values are lowest in the night, whereas $RH$ is generally very high at night on a tropical island like Singapore. 
These observations are supported by the time series plot in Fig.~\ref{fig:TimeSeries}. 

$PWV$ does not show strong correlation with any of the other features, except for a small positive correlation with $RH$. For temperate regions, a higher degree of correlation is observed between $RH$ and $PWV$ \cite{RH_PWV} as temperate regions have a much wider temperature range over different seasons and locations. This has a direct impact on the behavior and correlation of these variables.

Here the \emph{sine} and \emph{cosine} components of the $DoY$ and the $HoD$ features clearly show correlation with features like $T$, $RH$, $DPT$ and $SR$. 
The use of the cyclic properties of $HoD$ and $DoY$ help to clearly show the existing correlation between the weather variables as well as the seasonal and diurnal factors; this property was underestimated previously when the $HoD$ and the $DoY$ values were directly used instead of their \emph{sine} \& \emph{cosine} components~\cite{shilpa_IGARSS_feature}.

In summary, different weather features along with seasonal and diurnal features were identified. The correlations  between these features were studied and explained. However, only $RH$ and $T$ features have a very high correlation coefficient. This indicates that these weather variables can individually contribute towards a particular weather phenomenon. Therefore, in next section all these weather variables are studied with respect to rain events.

\subsection{Time Series Observation}
\label{sec:TimeSeries}
In this section, we study a time series plot of different weather features for the GPS station NTUS, shown in Fig.~\ref{fig:TimeSeries}. 

\begin{figure*}[tb]
\begin{center}
\subfloat[Rainfall \emph{classification}]
{\includegraphics[width=0.5\textwidth]{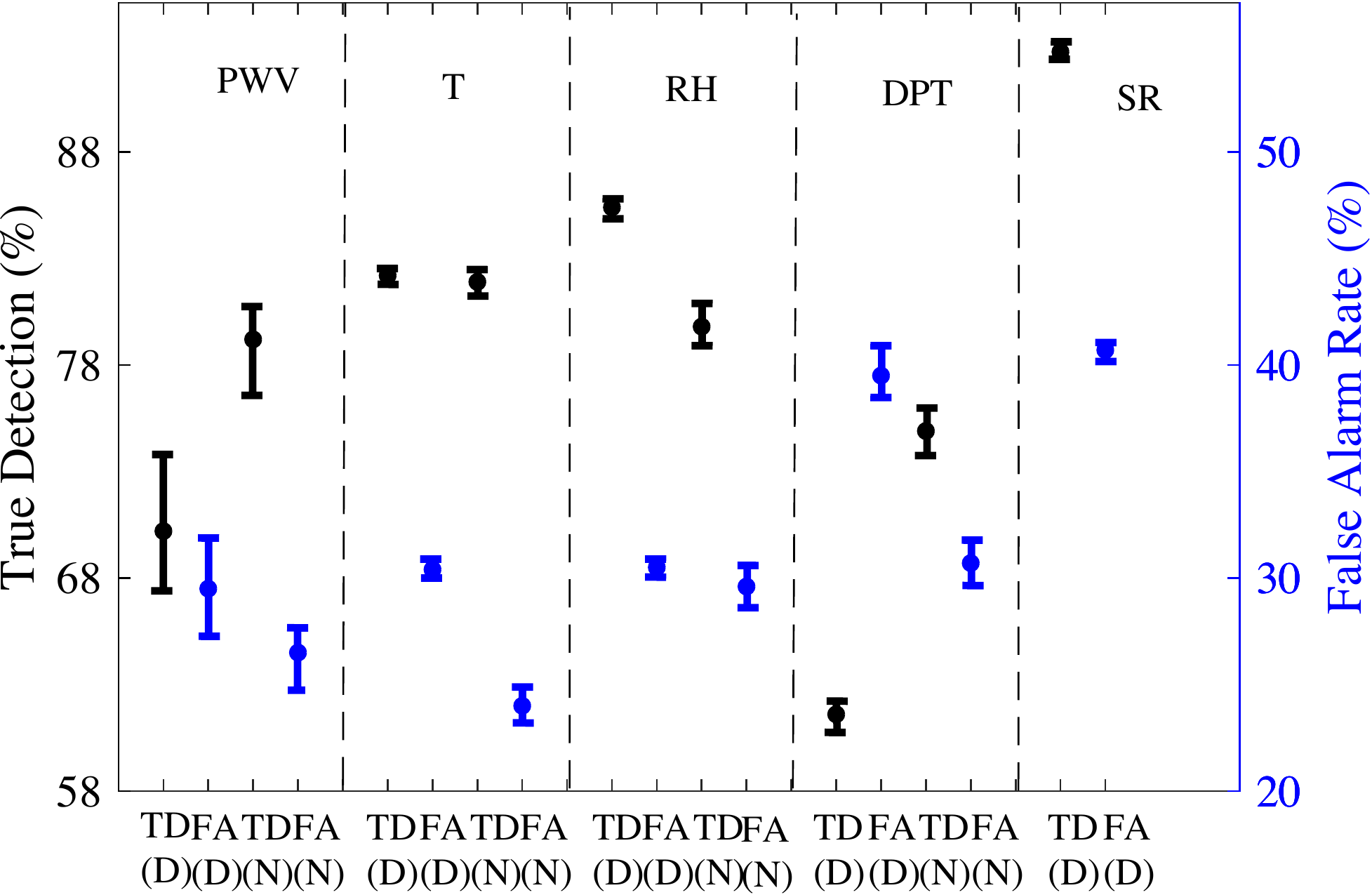}}
\subfloat[Rainfall \emph{prediction}]
{\includegraphics[width=0.5\textwidth]{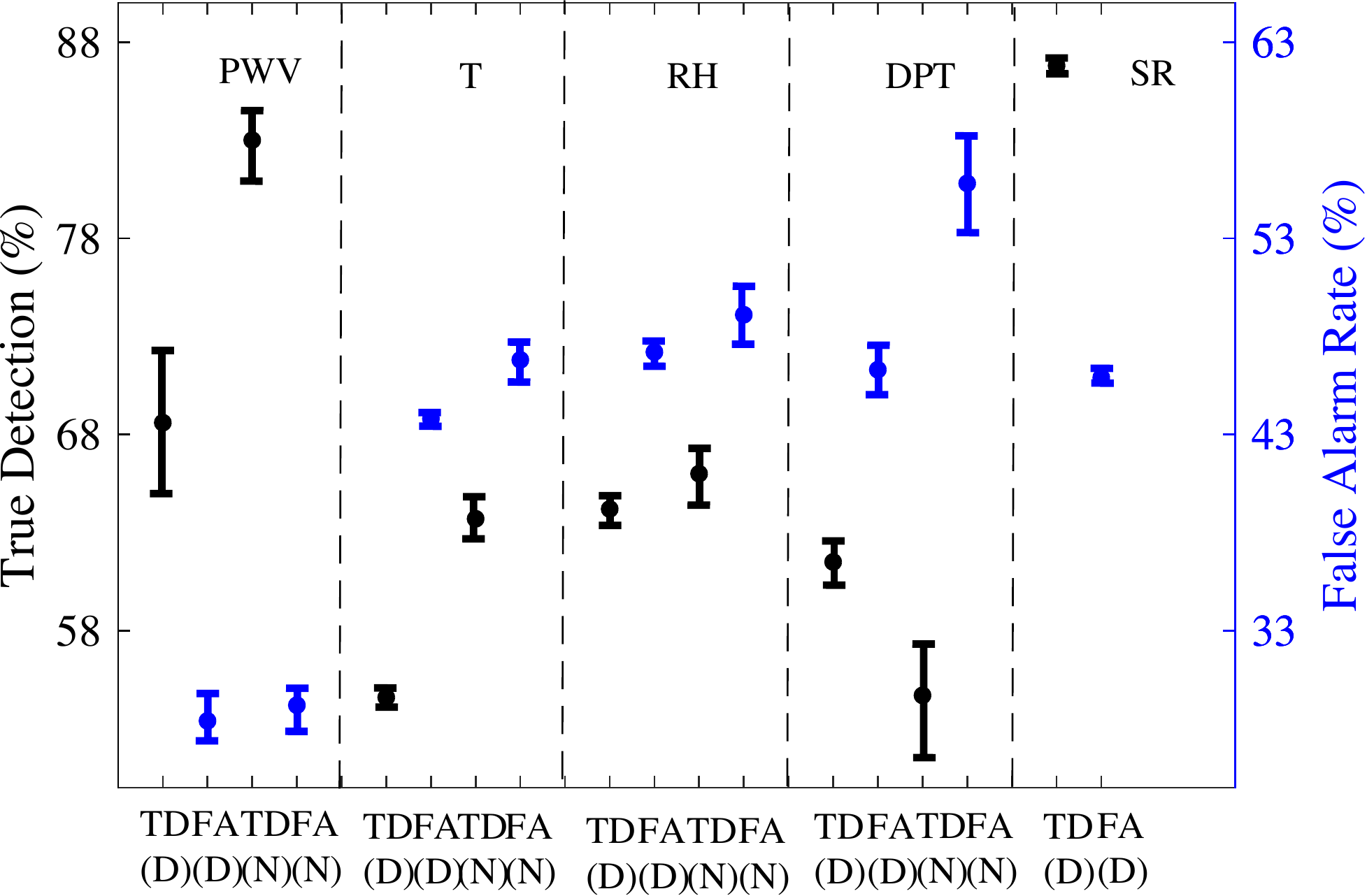}}
\caption{Performance of individual features for Rainfall \emph{classification} (a) and \emph{prediction} (b). $TD$ and $FA$ rates are represented in black and blue, respectively. (D) and (N) indicate results for Day and Night. Error bars show $95$\% confidence intervals. $TD$ ranges are shown on the left y axis, $FA$ ranges on the right (\emph{best viewed in color}).
\label{fig:bar_classify_predict}}
\end{center}
\end{figure*}

In window W1, rain events can be observed at 14:00 hours and at 18:00 hours. With respect to these rain events, different properties of the weather parameters can be analyzed. For both rain events, PWV significantly increases within $3$ hours before the rainfall starts. For the rain event at 14:00, the PWV values start to increase at around 11:00. For the rain event at 18:00, PWV starts to increase at around 15:00. Such an observation was also reported in \cite{TGRS_RainNowcasting}, which statistically showed that PWV in the tropical region increases within $2$ to $3$ hours before rainfall. 

Still in window W1, the temperature drops during the rain as the surface cools. It decreases and reaches values which are very similar to the dew point.  At the same time, a sharp increase in $RH$ can be noticed during the rain events, reaching $90-100$\%. The solar radiation on the other hand decreases before and during the rain due to the presence of the rain clouds. The WAHRSIS images taken during and before the rain events at 14:05  and at 17:30  respectively show the presence of thick dark clouds, which block the sun and lead to a significant decrease in solar radiation. 

Similar observations can be made from window W2, where a significant increase in PWV can be noticed before the start of the rain event.  PWV starts to increase at around 12:00  for a rain event that occurs at 15:30. The temperature drops to the dew point, and  $RH$  increases to $100$\% during the rain event. Similarly, the solar radiation  decreases before and during the rain event. The WAHRSIS image taken at 15:15 shows the presence of clouds. A clear sky image taken at 12:00 is also shown for reference (the black dot in the image marks the position of the Sun). Solar radiation can reach up to $1000$ W/m$^{2}$ in Singapore for a clear sky day  \cite{ClearSkyPIERS}, but   values can fall very low before and during a rain event.

Weather parameters like $T$, $RH$, $DPT$ and $SR$ show a very distinct behavior during the day and at night. These variables show fluctuations in the day time which could be correlated to rainfall, but in the night time they generally exhibit very little variation. As can be seen from Fig.~\ref{fig:TimeSeries}(b), relative humidity is always high (nearly $100$\%)  during the night and the early morning hours. Similarly, the temperature and the dew point readings are almost the same during the night and the early morning hours. Naturally, solar radiation is zero throughout the night. 

PWV also has a distinct diurnal pattern~\cite{Jin_diurnal,TGRS_RainNowcasting}, but unlike $T$, $DPT$, $RH$ and $SR$ values, PWV fluctuates during the night in response to rain events.  Window W3 shows a midnight rain event. A distinct increment in PWV can be observed $3$ hours before the start of the rain event at 00:30. This observation is correlated with the observations made for the day time rain events from sections W1 and W2. As expected,  $T$, $RH$, $DPT$ and the $SR$  do not show any significant changes corresponding to this midnight rain event.

In summary, we observed the different weather parameters which are important for rainfall prediction.  $T$, $RH$ and $DPT$ show sudden changes during the rain but not before. Both PWV and $SR$ show relatively distinct changes before a rain event in day time, but only PWV values show distinct changes before a night-time rain event. Time series plots for few more days are uploaded as supplemental material.

\section{Rainfall Prediction}
In our previous work \cite{TGRS_RainNowcasting} we used  $PWV$ and its second derivative to develop a model for the tropical region to predict a rain event with a lead time of $5$ minutes based on data from the $30$ minutes prior. The model was sub-divided into $3$ sections based on the seasons (NE-, SW- and Inter-monsoons). For this paper, we use the same rain prediction scenario whereby, (1) we divide our feature database into segments consisting of $30$ minutes of data, (2) for each $30$ minute segment, we check whether or not a rain event occurs after a lead time of $5$ minutes,  and (3) all rainfall within a $30$ minute window or less is considered as a single rain event~\cite{JX_1}. Different from \cite{TGRS_RainNowcasting}, here we study the combined effect of using different meteorological parameters along with $PWV$ in rainfall prediction. Instead of separate seasonal models, we combined them into a single model using the seasonal and diurnal features. 

The methodologies described in Section \ref{sec:method} are implemented to develop this rainfall prediction model. The evaluation metrics are reported after the model is trained and tested using data from years $2012$-$2015$ for NTUS station and data from year $2016$ for SNUS station.

\subsection{Assessment of Individual Features}

From the discussion of the time series observation (cf.\ Fig.~\ref{fig:TimeSeries}), it was observed that a few weather parameters ($T$, $RH$ and $DPT$) show sudden changes during rain compared to non-rain hours. Such a property is useful in classifying rain and no-rain conditions. However, since these weather variables do not show any significant changes before the start of a rain event, they might not be useful for rainfall prediction. On the other hand, weather variables like $PWV$ and $SR$ do show significant changes before the start of a rain event, which are useful for rainfall prediction. 
Thus, in this section, we analyze the performance of  individual features for rainfall classification and prediction. To show the effect of the time-of-day, the results are segregated into day and night time.

The results for this section are obtained by using four years ($2012$-$2015$) of data from the NTUS station. Fig.~\ref{fig:bar_classify_predict}(a) shows the rainfall \emph{classification} results in terms of true detection and false alarm rates for day and night. Similarly Fig.~\ref{fig:bar_classify_predict}(b) shows the results for rainfall \emph{prediction}. From Fig.~\ref{fig:bar_classify_predict}(a), it can be observed that all features can clearly differentiate rain and no-rain conditions in the day time. Most of these features have good performance for rainfall classification in night-time as well except for solar radiation. Since there is no solar radiation at night, it has no effect on the rain classification or prediction; therefore, we report only day time results for $SR$.

From Fig.~\ref{fig:bar_classify_predict}(b) it can be observed that unlike in the classification scenario, not all  features  have good true detection and false alarm rates for rainfall \emph{prediction}. Similar to the rainfall classification results, $SR$ gives the highest $TD$ during daytime, whilst it is not useful in the night. While features like $T$, $RH$ and $DPT$ show a good capability in rainfall \emph{classification}, they are not very useful for rainfall \emph{prediction} as these parameters change only during the rain but not before. $PWV$ is the only feature which shows a good separation between $TD$ and $FA$ for rainfall prediction in both day and night cases. These results are consistent with our time series observation in Fig.~\ref{fig:TimeSeries}.

Therefore, it is clear that not all features are useful for rainfall prediction, and the accuracy is expected to improve with the inclusion of the diurnal and seasonal features combined with the weather features. 

\begin{figure*}[tb]
\begin{center}
\subfloat[Adding features]
{\label{fig:addition}\includegraphics[width=0.5\textwidth]{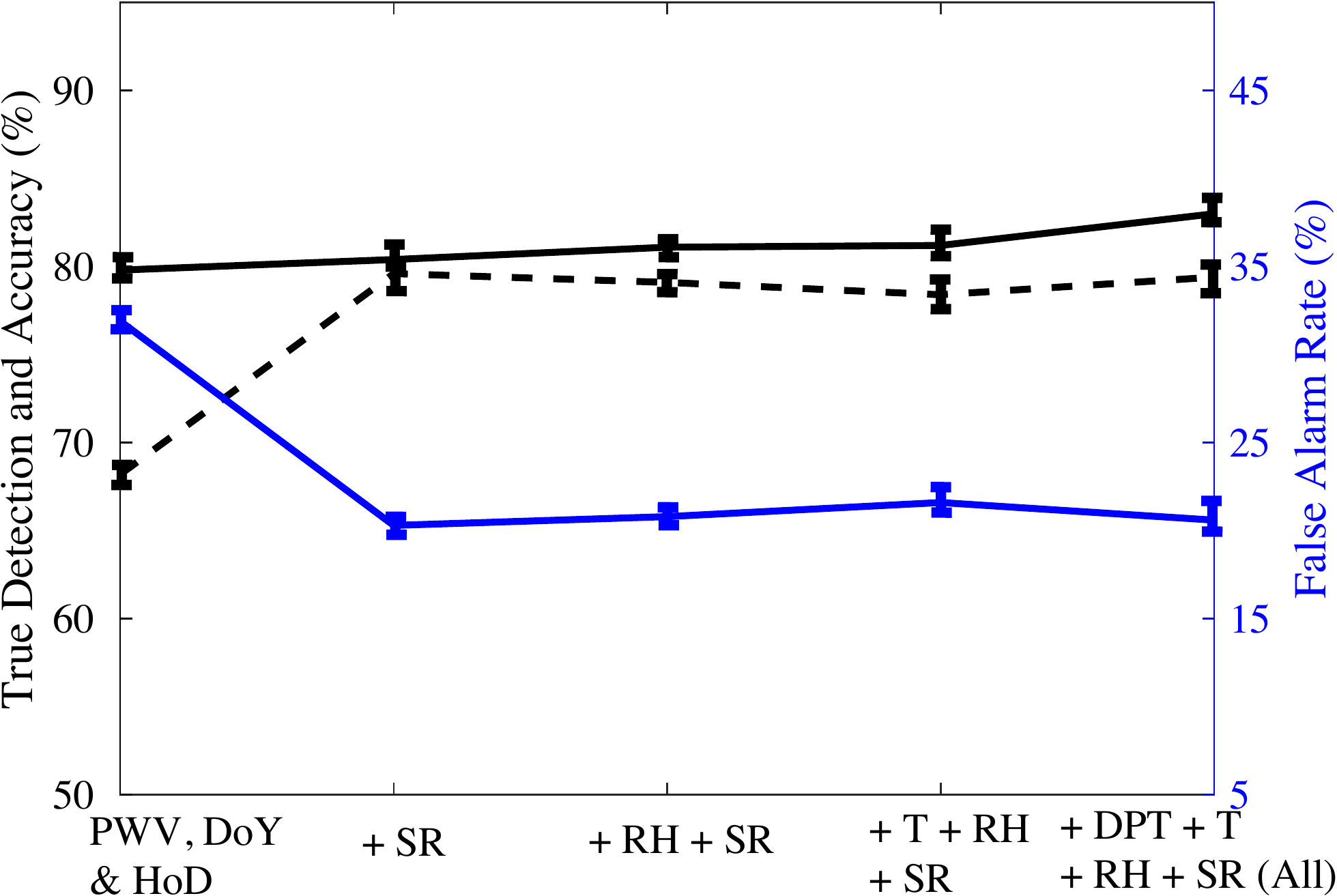}}
\subfloat[Removing features]
{\label{fig:eleminate}\includegraphics[width=0.5\textwidth]{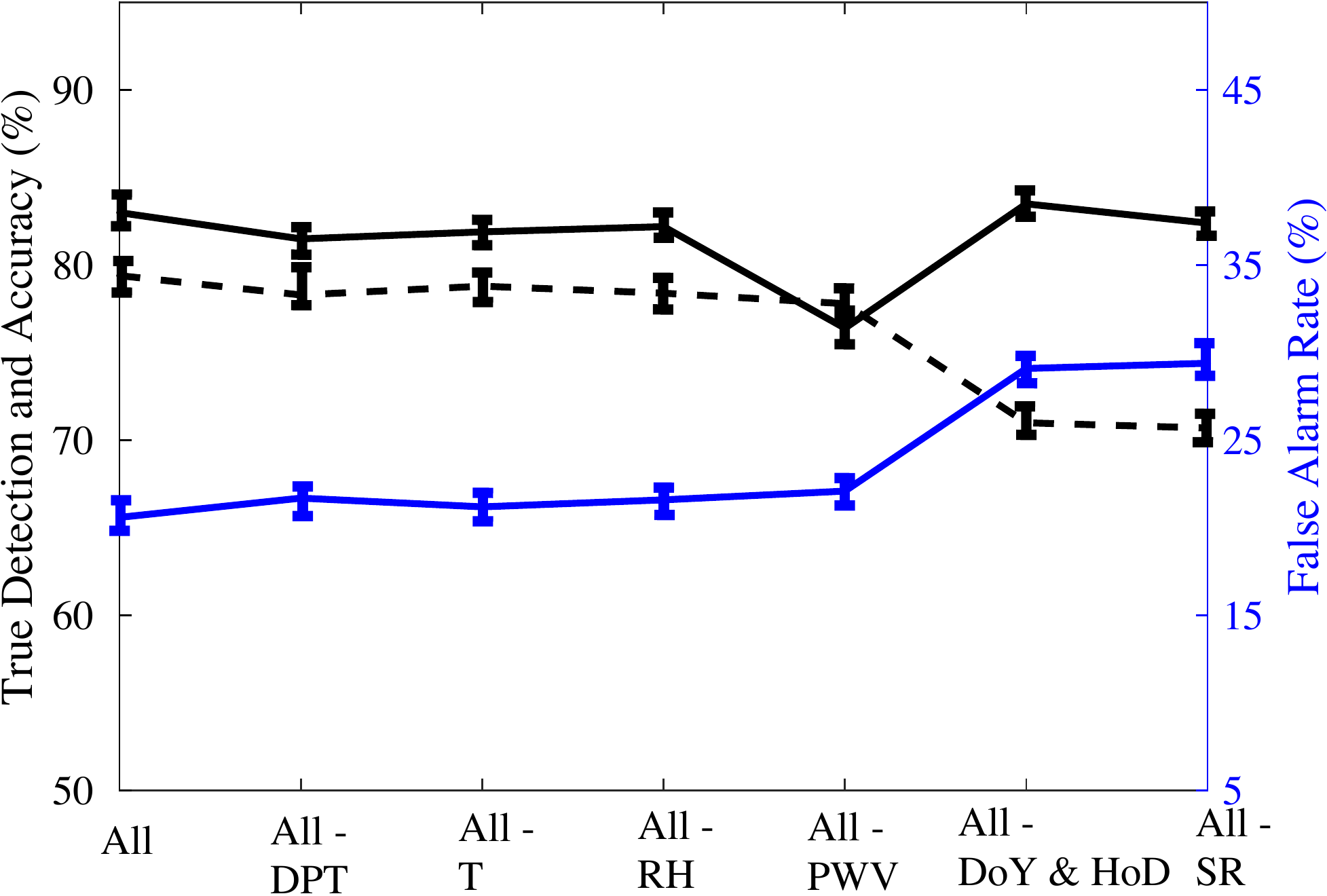}}
\caption{Rainfall prediction performance  when different features are (a) added or (b) removed one at a time. $TD$ and $FA$ rates are represented by solid black and blue lines, respectively, while $A$ is represented by a dashed black line.  Error bars show $95$\% confidence intervals. $TD$ and $A$ ranges are shown on the left y axis, $FA$ ranges on the right (\emph{best viewed in color}).
\label{fig:adding-removing}}
\end{center}
\end{figure*}

\subsection{Selection of Optimal Features}

In \cite{TGRS_RainNowcasting}, rainfall prediction is done based on PWV and seasonal behavior only. In this section, we analyze how the rainfall prediction performance improves or deteriorates by adding or removing specific features. The results discussed in the following are obtained by using the four years ($2012$-$2015$) data from the NTUS station. 

\subsubsection{Adding Features} 
We start by taking the combination of $PWV$, $HoD$ and $DoY$ features as suggested above. Then the other features are added successively. The first reading in Fig.~\ref{fig:addition} is obtained by using the combination of $PWV$,  seasonal ($DoY$) and diurnal ($HoD$) features. The subsequent readings are obtained when other features are added to the previous pool as shown by the labels. We run the experiment over 50 iterations, with $30$\% of the total data as the training set and remainder as the test set. The reported evaluation metrics are an average over the $50$ iterations. Therefore, along with the mean evaluation metrics, we also report their $95$\% confidence intervals.

The first reading of Fig.~\ref{fig:addition} shows a $TD$ rate of $79.8$\%, $FA$ rate of $31.9$\%, and an overall accuracy of $68.2$\% for rainfall prediction with a lead time of $5$ minutes. When $SR$ is added in the second step, the $FA$ rate decreases significantly to $20.3$\%, the $TD$ rate improves and reaches $80.4$\%. Therefore, the overall accuracy increases to $79.6$\%. 

From Fig.~\ref{fig:bar_classify_predict} it was observed that the features $T$, $RH$ and $DPT$ are not significant in rainfall prediction. Therefore, when these features are subsequently added to the pool of features, $TD$ increases but so does $FA$.  
When all the features are involved, the $TD$, $FA$ and $A$ values are $83.0$\%, $20.6$\% and $79.4$\% respectively. This experiment shows that the highest accuracy is achieved when the feature combination of $PWV$, $SR$, $DoY$ and $HoD$ is used.

\subsubsection{Removing Features} 
To further elaborate on the above findings, we now systematically eliminate features from the pool one at a time and analyze prediction performance, as shown in Fig.~\ref{fig:eleminate}. The first reading is obtained by using all  $7$ features for the rainfall prediction. The subsequent readings are when the respective feature as shown by the x-axis of the plot are removed from the pool of all the $7$ features leaving behind always $6$ features. When only $RH$ is removed from the pool, it is labeled ``All - RH''. For our analysis, the readings corresponding to the elimination of an individual feature are compared to the first reading when all the features are used.

As mentioned earlier, when all the features are used, a $TD$ value of $83.0$\%, a $FA$ value of $20.6$\% and an overall accuracy of $79.4$\% are obtained. When one of the features $DPT$, $T$, or $RH$ is removed from the pool of all features, $TD$ values decrease slightly and $FA$ and $A$ values remain same as that of the first reading. This indicates that the presence of either of these features do not contribute much for rainfall prediction. However, when the $PWV$ feature is eliminated, a significant drop in the $TD$ values can be noticed as compared to the first reading. This indicates that the $PWV$ values play an important role in maintaining a high detection rate and high accuracy. When the seasonal and the diurnal features ($DoY$ and $HoD$) are removed from the pool of all the features, $TD$ remains almost the same, whereas $FA$ increases by almost $8.5$\% and the accuracy decreases compared to the first reading. Similarly, when the $SR$ feature is eliminated, the $FA$ values increase from $20.6$\% to $29.4$\%. These results show that the features $SR$, $DoY$ and $HoD$ are important in helping to reduce the false alarm rate and increase the accuracy. 

Therefore, the features $DPT$, $RH$ and $T$ do not contribute to improving the $TD$ rates, and removal of these features one at a time actually improves the accuracy by slightly reducing false alarms. On the other hand, $HoD$, $DoY$ and $SR$ are important to control the false alarms, and $PWV$ is important to maintain a high $TD$ rate. This confirms our previous work \cite{TGRS_RainNowcasting}, where $PWV$ is shown to provide a good rainfall prediction ability. Similarly, different models were proposed for different seasons for lowering the $FA$ rates in \cite{TGRS_RainNowcasting}, which is in line with the newly introduced $DoY$ and $HoD$ features and their importance for $FA$ reduction.

In summary, we conclude from these experiments that the combination of the features $PWV$, $SR$, $DoY$ and $HoD$ gives the best results. The $TD$ rate for this combination is $80.4$\%, which corresponds to a missed detection rate  of around $19.6$\%. The $FA$ rate for this combination is $20.3$\%, and the overall accuracy is $79.6$\%.

\subsection{Benchmarking}
In this section, the results obtained by using $PWV$, $SR$, $HoD$ and $DoY$ features for rainfall prediction are compared to the literature. The true detection and false alarm rates achieved by our proposed approach  show a significant improvement -- especially in terms of false alarm rates -- over those reported in \cite{Benevides,Yibin(a),Yao_2018}, see Table \ref{table:benchmarking0}.  

\begin{table}[htb]
\begin{center}
\caption{Benchmarking with other approaches.}
\begin{tabular}{|r|c|c|}
\hline
Approach  & \shortstack{TD (\%)} &  \shortstack{FA (\%)} \\
\hline
Proposed & 75-88 & 19-23 \\
\cite{Benevides} & 75 & 60-70 \\
\cite{Yibin(a)}  & 80 & 66 \\
\cite{Yao_2018}  & 85 & 66 \\
\hline
\end{tabular}
\label{table:benchmarking0}
\end{center}
\end{table}

Table~\ref{table:benchmarking} shows a detailed comparison  between the proposed method and the results reported in \cite{TGRS_RainNowcasting}. For all years ($2012$-$2015$), by incorporating the effect of $SR$, $HoD$ and $DoY$ with the $PWV$ feature, the false alarm rate can be significantly lowered. On average, the $FA$ rate is reduced by $16.2$ percentage points, with only a small reduction in $TD$. Similar results are observed for the PWV data from the SNUS GPS station and the ground based data from the co-located weather station: When the features $PWV$, $SR$, $DoY$ and $HoD$ are used for the rainfall prediction, the $FA$ rates are reduced by to less than half. 

\begin{table}[htb]
\begin{center}
\caption{Comparison with \cite{TGRS_RainNowcasting}.}
\begin{tabular}{|c | c |c |c |c |c|}
\hline 
Station & \shortstack{Years \\ (Rain Events)} & \multicolumn{2}{c|}{\shortstack{ Literature \cite{TGRS_RainNowcasting}\\ ($PWV$)}} & \multicolumn{2}{c|}{\shortstack{Applied Algorithm \\ ($PWV$, $SR$,  \\ $DoY$ \& $HoD$)}} \\
\hline
{} & {}  & \shortstack{TD (\%)} &  \shortstack{FA (\%)}  & \shortstack{TD (\%)} &  \shortstack{FA (\%)}  \\
\hline
NTUS & {2012 (219)} &  88.5   & 36.4   & 75.0  & 19.1 \\
	 & {2013 (252)} &  84.9   & 43.7   & 78.8  & 23.1 \\
     & {2014 (231)} &  89.1   & 34.6   & 82.3  & 18.8 \\
     & {2015 (222)} &  89.1   & 31.0   & 87.9  & 19.9 \\
     & Average      &  87.9   & 36.4   & 81.0  & 20.2  \\
\hline
SNUS & {2016 (195)} &  90.7   & 50.2   & 81.1  & 19.2\\
\hline
\end{tabular}
\label{table:benchmarking}
\end{center}
\end{table}

\section{Conclusions \& Future Work}
\label{sec:conc}
We have identified the different ground based weather features which are important for the prediction of rain events. A detailed analysis is done to study the interdependency of these variables. We have incorporated seasonal and diurnal factors into the model, along with weather variables. All the features play a significant role in rainfall classification, while features like $PWV$, $SR$, $DoY$ and $HoD$ in particular show potential for rainfall prediction as well. The $PWV$ feature contributes the most to achieving a high detection rate, and the features $SR$, $DoY$ and $HoD$ contribute to a reduction of false alarm rates. Compared to the literature, our approach achieves a significant reduction of $FA$ rates.  

As future work, we plan to study the impact of using different weather features for different stations with a larger dataset.
We will also consider the derivatives of features, as well as additional features like wind, cloud coverage, etc.\ for rainfall prediction with longer lead times.

\bibliographystyle{IEEEbib}


\begin{IEEEbiography}[{\includegraphics[width=1in,height=1.25in,clip,keepaspectratio]{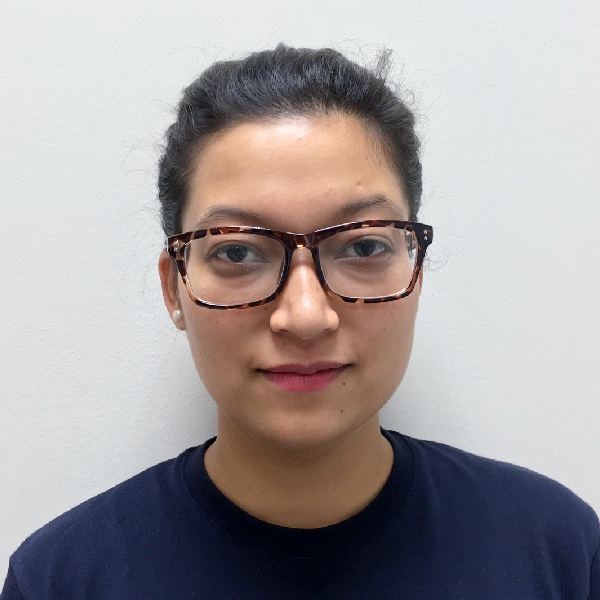}}]{Shilpa Manandhar} (S\textquotesingle14) received the B.Eng.\ degree (Hons.) in Electrical and Electronics engineering from Kathmandu University, Dhulikhel,
Nepal, in 2013, and the PhD degree from Nanayang Technological University (NTU), Singapore, in 2019. She is currently working as a Research Fellow in NTU.
Her research interests include remote sensing, study of global positioning system signals to predict meteorological phenomenon
such as rain and clouds, and improvement of GPS localization.
\end{IEEEbiography}

\begin{IEEEbiography}
[{\includegraphics[width=1in,height=1.25in,clip,keepaspectratio]{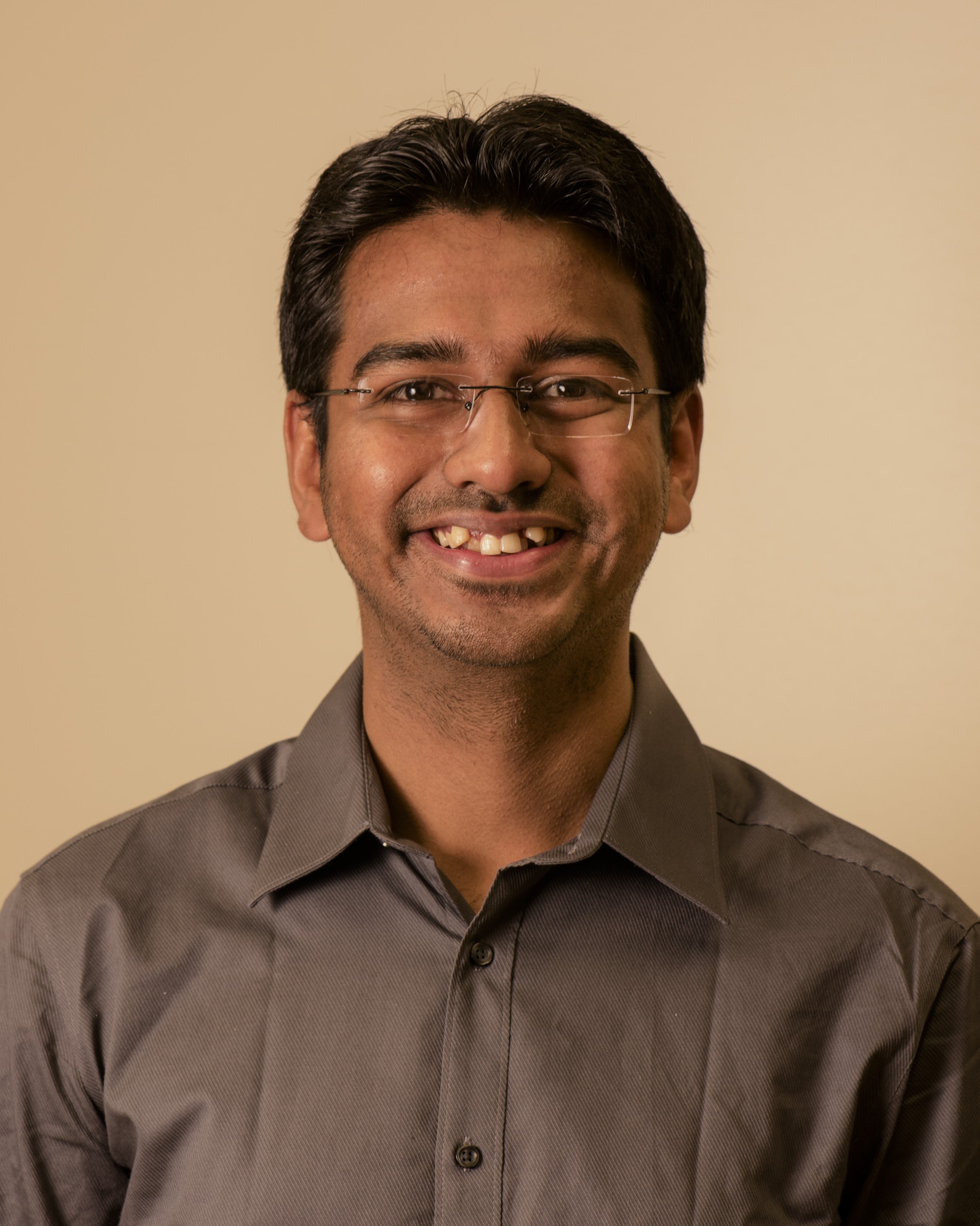}}]{Soumyabrata Dev} (S\textquotesingle09-M\textquotesingle17) graduated summa cum laude from National Institute of Technology Silchar, India with a B.Tech.\ in 2010. Subsequently, he worked in Ericsson  as a network engineer from 2010 to 2012. Post that, he obtained his Ph.D.\ from Nanyang Technological University (NTU) Singapore, in 2017. From Aug-Dec 2015, he was a visiting student at Audiovisual Communication Laboratory (LCAV), \'{E}cole Polytechnique F\'{e}d\'{e}rale de Lausanne (EPFL), Switzerland. Currently, he is a Postdoctoral Researcher at ADAPT Centre, Dublin, Ireland. His research interests include remote sensing, statistical image processing, machine learning, and deep learning.
\end{IEEEbiography}

\begin{IEEEbiography}[{\includegraphics[width=1in,height=1.25in,clip,keepaspectratio]{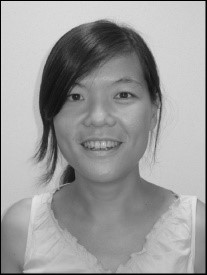}}]{Yee Hui Lee}
(S\textquotesingle96$-$M\textquotesingle02$-$SM\textquotesingle11) received the B.Eng. (Hons.) and M.Eng. degrees from the School of Electrical and Electronics Engineering, Nanyang Technological University, Singapore, in 1996 and 1998, respectively, and the Ph.D. degree from the University of York, York, U.K., in 2002. She is currently an Associate Professor with the School of Electrical and Electronic Engineering, Nanyang Technological University, where she has been a Faculty Member since 2002. Her research interests include channel characterization, rain propagation, antenna design, electromagnetic bandgap structures, and evolutionary techniques.
\end{IEEEbiography}

\begin{IEEEbiography}[{\includegraphics[width=1in,height=1.25in,clip,keepaspectratio]{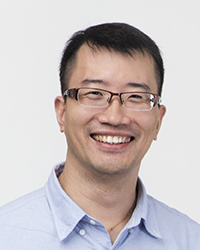}}]{Yu Song Meng}
(S\textquotesingle09$-$M\textquotesingle11) received the B.Eng. (Hons.) and Ph.D. degrees in electrical and electronic engineering from Nanyang Technological University, Singapore, in 2005 and 2010 respectively.

He was a Research Engineer with the School of Electrical and Electronic Engineering, Nanyang Technological University, from 2008 to 2009. He joined the Institute for Infocomm Research, Agency for Science, Technology and Research (A*STAR), Singapore, in 2009 as a Research Fellow and then a Scientist I. In 2011, he was transferred to the National Metrology Centre, A*STAR, where he is currently appointed as a Senior Scientist I. From 2012 to 2014, he was part-timely seconded to Psiber Data Pte. Ltd., Singapore, where he was involved in metrological development and assurance of a handheld cable analyser, under a national Technology for Enterprise Capability Upgrading (T-Up) scheme of Singapore. Concurrently, he also serves as a Technical Assessor for the Singapore Accreditation Council-Singapore Laboratory Accreditation Scheme (SAC-SINGLAS) in the field of RF and microwave metrology. His current research interests include electromagnetic metrology, electromagnetic measurements and standards, and electromagnetic-wave propagations.

Dr. Meng is a member of the IEEE Microwave Theory and Techniques Society. He is a recipient of the Asia Pacific Metrology Programme (APMP) Iizuka Young Metrologist Award in 2017 and the national T-Up Excellence Award in 2015. 
\end{IEEEbiography}

\begin{IEEEbiography}[{\includegraphics[width=1in,height=1.25in,clip,keepaspectratio]{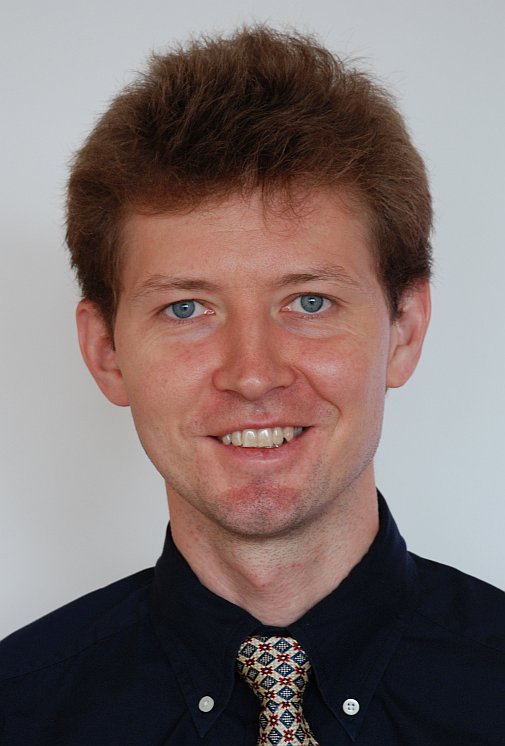}}]{Stefan Winkler} (FS\textquotesingle18--SMS\textquotesingle14) is Deputy Director at AI Singapore and Associate Professor (Practice) at the National University of Singapore. Prior to that, he was Distinguished Scientist and Director of the Video \& Analytics Program at the University of Illinois' Advanced Digital Sciences Center (ADSC) in Singapore. He also co-founded two start-ups (Opsis and Genista) and worked for a Silicon Valley company (Symmetricom). 

Dr.\ Winkler has a Ph.D.\ degree from the Ecole Polytechnique F\'{e}d\'{e}rale de Lausanne (EPFL), Switzerland, and a Dipl.-Ing.\ (M.Eng./B.Eng.) degree from the University of Technology Vienna, Austria.  He is an IEEE Fellow and has published over 130 papers.  His research interests include video processing, computer vision, machine learning, perception, and human-computer interaction.

\end{IEEEbiography}
\end{document}